\documentclass[conference]{IEEEtran}

\IEEEoverridecommandlockouts

\usepackage[utf8]{inputenc}

\usepackage{tabularray}

\usepackage{csquotes}

\usepackage[numbers]{natbib}

\usepackage{graphicx}
\graphicspath{{images/}}

\usepackage{amsmath}

\usepackage{eurosym}

\usepackage[hyphens]{url}
\usepackage{hyperref}
\usepackage[hyphenbreaks]{breakurl}

\usepackage{array}

\usepackage{multirow}

\usepackage{makecell}

\usepackage{float}
\usepackage{dblfloatfix}
\usepackage{tcolorbox}

\usepackage{graphicx} 
\usepackage{subcaption}

\begin{document}

\title{Balancing Progress and Responsibility: A Synthesis of Sustainability Trade-Offs of AI-Based Systems}

\author{
    \IEEEauthorblockN{
        Apoorva Nalini Pradeep Kumar\IEEEauthorrefmark{1}, Justus Bogner\IEEEauthorrefmark{1}, Markus Funke\IEEEauthorrefmark{1}, Patricia Lago\IEEEauthorrefmark{1}
    }
    \IEEEauthorblockA{
        \IEEEauthorrefmark{1}Vrije Universiteit Amsterdam, Amsterdam, The Netherlands\\
        a.nalini.pradeep.kumar@student.vu.nl, j.bogner@vu.nl, m.t.funke@vu.nl, p.lago@vu.nl
    }
}

\maketitle


\begin{abstract}
Recent advances in artificial intelligence (AI) capabilities have increased the eagerness of companies to integrate AI into software systems.
While AI can be used to have a positive impact on several dimensions of sustainability, this is often overshadowed by its potential negative influence.
While many studies have explored sustainability factors in isolation, there is insufficient holistic coverage of potential sustainability benefits or costs that practitioners need to consider during decision-making for AI adoption.
We therefore aim to synthesize trade-offs related to sustainability in the context of integrating AI into software systems.
We want to make the sustainability benefits and costs of integrating AI more transparent and accessible for practitioners.

The study was conducted in collaboration with a Dutch financial organization.
We first performed a rapid review that led to the inclusion of 151 research papers.
Afterward, we conducted six semi-structured interviews to enrich the data with industry perspectives.
The combined results showcase the potential sustainability benefits and costs of integrating AI.
The labels synthesized from the review regarding potential sustainability benefits were clustered into 16 themes, with \textit{energy management} being the most frequently mentioned one.
11 themes were identified in the interviews, with the top mentioned theme being \textit{employee wellbeing}.
Regarding sustainability costs, the review discovered seven themes, with \textit{deployment issues} being the most popular one, followed by \textit{ethics \& society}.
\textit{Environmental issues} was the top theme from the interviews.
Our results provide valuable insights to organizations and practitioners for understanding the potential sustainability implications of adopting AI.
\end{abstract}

\begin{IEEEkeywords}
artificial intelligence, sustainability, trade-offs, costs, benefits, rapid review, interviews
\end{IEEEkeywords}

\section{Introduction}
The role of artificial intelligence (AI) in our everyday lives has tremendously increased in the past few years. AI and its sub-fields like machine learning (ML) have gained attention due to their enhanced capabilities to provide functionalities in software systems that were previously not possible~\cite{Jordan2015Machine}. Gartner suggests that more than 80\% of enterprises will have used Generative AI APIs or deployed generative AI-enabled applications by 2026.\footnote{\url{https://www.gartner.com/en/newsroom/press-releases/2023-10-11-gartner-says-more-than-80-percent-of-enterprises-will-have-used-generative-ai-apis-or-deployed-generative-ai-enabled-applications-by-2026}}
Over the years, smart, intelligent and adaptive software applications have been built by integrating them with AI. Examples of these applications include recommendation systems in e-commerce applications and auto-completion in word-processing systems. A market analysis report posits that the global AI market size was valued at USD 136.55 billion in 2022 and is projected to expand at a compound annual growth rate of 37.3\% from 2023 to 2030.\footnote{\url{https://www.grandviewresearch.com/industry-analysis/artificial-intelligence-ai-market}}

An example of a potential sustainability-related benefit of AI adoption is reducing the repetitive work of employees, while a potential sustainability-related cost could be the heightened energy consumption for training AI models. Considering and prioritizing the sustainability implications of AI in decision-making is gaining importance for organizations to ensure compliance with new regulatory standards and to maintain their brand reputation concerning privacy and data security. Given the energy-hungry nature of AI, the decision to adopt AI or not is contingent on being aligned with the organizations' larger sustainability commitments.\footnote{\url{https://www.europarl.europa.eu/news/en/headlines/society/20230601STO93804/eu-ai-act-first-regulation-on-artificial-intelligence}} Therefore, organizations need to assess the sustainability implications when incorporating or employing AI to ensure they minimize adverse impacts on sustainability while also meeting their sustainability objectives. To allow companies to adopt AI sustainably and responsibly, this study focuses on investigating and synthesizing the potential sustainability-related benefits and costs of integrating AI into software systems. We partnered with a large Dutch financial organization and conducted a rapid review to synthesize the sustainability benefits and costs of AI integration software systems. Afterwards, we conducted multiple semi-structured interviews with practitioners from the same company plus external consultants to investigate the trade-offs of AI integration from an industry perspective.
Although multiple studies have been published on the benefits and harms of the use of AI, this study aims to aggregate valuable insights for practitioners regarding the possible sustainability trade-offs in integrating AI, and thereby facilitates increased awareness.

\section{Background and Related Work}

\subsection{Software Sustainability}
The Brundtland Report~\cite{Keeble1988} defines Sustainable Development as \enquote{the development that meets the needs of the present without compromising the ability of future generations to meet their own needs}. 
Consequently, Lago et al.~\cite{Lago2015} defined sustainability as a property of software quality and categorized sustainability concerns regarding software-intensive systems into four key dimensions: \textit{environmental}, \textit{social}, \textit{economic}, and \textit{technical}. The environmental dimension emphasizes the preservation of natural resources while improving human welfare, e.g., addressing energy efficiency and ecological requirements. The social dimension focuses on ensuring equitable access to social resources for the current and future generations. For software-intensive systems, it is aimed at directly or indirectly supporting social communities in any domain and mitigating negative impacts of software on society. Preserving capital and economic value is the aim of economic sustainability, e.g., ensuring that the development and provisioning of software stays affordable over the years. Lastly, the focus of technical sustainability is on the long-term use of software-intensive systems and their appropriate evolution in an execution environment that continuously changes. Hence, as the adoption of AI in various sectors continues to grow, it becomes increasingly important to not overlook the sustainability implications of this trend.\footnote{\url{https://openai.com/research/ai-and-compute}}

\subsection{Related Work}
The recent advancements in AI promise to revolutionize different fields like agriculture, e-commerce and finance through increased adoption~\cite{Javaid2023Understanding,Pallathadka2023Applications}. Although many studies focus on applications of AI that lead to improvements in their corresponding fields, many of these papers are not focused on the direct sustainability implications of such applications.

In a position paper, Vinuesa et al.~\cite{Vinuesa2020role} detailed how AI could help in achieving 79\% of the targets of the Sustainable Development Goals (SDGs) while also hindering 35\% of the targets by classifying the SDGs based on economic, environmental and social outcomes. Further, Gupta et al.~\cite{Gupta2021Assessing} conducted a panel discussion that highlighted the potential of AI in solving existing problems related to sustainability as well as the shortcomings of AI that would result in negative outcomes for society. The study also presses on the importance of the potential positive and negative impact of technology not just in the particular domain of deployment but across the general environment, economy and society.

In their study, Galaz et al.~\cite{Galaz2021Artificial} delved into the various sustainability risks and negative impacts of AI. They categorize them into algorithmic bias and allocative harms, unequal access and benefits, cascading failures and external disruptions, and trade-offs between efficiency and resilience from a review of the literature. Characterising the harms in the machine learning cycle, Suresh and Guttag~\cite{UnderstandingHarm} laid out seven potential sources of bias such as historical, representation bias, measurement bias etc., and discussed mitigation strategies for each. 

Schwartz et al.~\cite{GreenAISchwartz} talk about Red AI that focuses on achieving high levels of accuracy without caring about energy usage. The paper also points out diminishing returns, i.e., a linear increase in accuracy is attained with an exponential increase in data to be trained. Inversely, Brownlee et al.~\cite{AccuracyEnergyTradeoff} carried out experiments to show the accuracy-energy trade-off in machine learning and how a small reduction in accuracy can lead to a significant saving of energy in some cases.

The recent paper by Moises de Souza~\cite{Moises2023Social} provides a systematic review of approaches to include social sustainability factors in the general software development lifecycle, not only for AI-based systems. Particularly, some supporting techniques were identified that could tackle social sustainability in software development. Raising awareness among practitioners was the primary technique discussed in the review, as it would be the first step in building software that is sustainable along all dimensions.

Many of the aforementioned studies focus on specific implications and considerations of AI for \textit{individual} dimensions of sustainability. However, there is so far no comprehensive analysis of the benefits and costs of integrating AI for \textit{all} sustainability dimensions. Such an analysis could support practitioners in the decision-making about AI adoption. The paper thus aims to bridge this identified gap by conceptualizing the benefits and costs of AI integration for increased awareness, transparency, and usability in decision-making concerning AI-based systems.

\section{Study Design}
\label{sec:design}
In this section, we describe the detailed design of our study. The general goal of the research is stated as follows:

\begin{center}
\textit{Analyze the potential costs and benefits} \\
\textit{For the purpose of knowledge generation and awareness} \\
\textit{With respect to sustainability} \\
\textit{From the point of view of software practitioners} \\
\textit{In the context of AI adoption in companies}
\end{center}

This research objective is divided into two research questions that each address one of the two concepts.\\
\textbf{RQ1:} What are the potential sustainability \textbf{benefits} of integrating AI into software systems? \\
\textbf{RQ2:} What are the potential sustainability \textbf{costs} associated with integrating AI into software systems?\\
To reach this goal and answer these RQs, we partnered with a large Dutch financial organization of 10,000+ employees.
This bank actively tries to incorporate a sustainability perspective into their IT and software development teams.
They have 140+ software architects with responsibility for their applications, who regularly have to make design decisions related to sustainability.
First, we conducted a rapid review~\cite{Cartaxo2020Rapid} of relevant literature to analyse the sustainability benefits and costs of integrating or using AI in software systems. These results were presented to our industry partner. To enrich and extend these results from scientific literature with practitioner opinions and experiences, we followed up with semi-structured interviews~\cite{HoveExperiences}, both with employees of the financial organization and external consultants of theirs. Afterwards, we performed thematic analysis~\cite{Cruzes2011} for the findings from both and analysed differences and commonalities.
Lastly, we compiled the final results into an evidence briefing.
For transparency and reusability, important artifacts and documents related to the rapid review and the interviews are available via Zenodo.\footnote{\url{https://www.doi.org/10.5281/zenodo.10523911}}

\subsection{Rapid Review} 
A rapid review is a lightweight secondary study that aims to efficiently identify and analyse the existing literature on an industry-relevant topic in close collaboration with practitioners~\cite{Cartaxo2020Rapid}. We selected this research method to get high-quality results from the literature in a short time frame and its suitability for academia-industry collaboration. The outcome of this was a thematically classified list of potential sustainability-related benefits and costs of integrating AI into software systems. The protocol for the rapid review is outlined below.
\subsubsection{Search Strategy} The search strategy for finding literature was selected to minimize information loss while best answering our research questions. After trying out many different search angles, we finally selected two search terms with keywords that yielded a high ratio of true positives:

\begin{itemize}
    \item (AI OR \enquote{artificial intelligence} OR ML OR \enquote{machine learning}) (sustainable OR sustainability OR tradeoff OR trade-off)
    \item AI OR \enquote{artificial intelligence} OR ML OR \enquote{machine learning}) (cost OR benefit OR downside OR danger OR harm OR negative OR positive OR impact OR effect OR risk)
\end{itemize}

Google Scholar was selected as the search engine for its comprehensive coverage, with the search being performed only for the title field. For efficiency, a stopping criterion was established to only look at the first 250 results of each search query, i.e., we considered 500 publications for inclusion. As Google Scholar sorts its results by their relevance to the search terms, this strategy was selected to find the most relevant papers for the study. The first author filtered this initial set based on the inclusion and exclusion criteria. For all included papers, the first author then extracted the relevant data to answer the RQs. During selection and extraction, regular discussions to clarify non-trivial decisions and establish consistency took place. After the first round of data extraction, papers that discussed the broader implications of AI concerning sustainability rather than those concentrating on specific applications were selected for one round of backward and forward snowballing. The scope of the snowballing was limited to 30 references and citations for each paper meeting this criterion. This \enquote{focused snowballing} increased our coverage of publications in exchange for manageable efforts. 
\subsubsection{Study Selection}
The following inclusion and exclusion criteria were used to select or reject the publications. A publication could be included if it satisfied both of these criteria:

\begin{itemize}
    \item Article focuses on potential costs, benefits or trade-offs associated with integrating or adopting AI
    \item The costs, benefits or trade-offs discussed in the article are (at least partly) related to sustainability
\end{itemize}

Regardless of the two inclusion criteria, a publication was rejected if it satisfied one of the following:

\begin{itemize}
    \item Article is a non-English publication
    \item Full text of the published article not available
    \item Article is not a scientific paper (but: peer review not required)
    \item Article is a duplicate or extension of an included study
    \item Secondary or tertiary studies
\end{itemize}

\subsubsection{Data Extraction and Synthesis}
\label{para:dataexsyn}
Data relating to sustainability costs and benefits discussed in each of the selected papers was extracted and exported to a spreadsheet. Metadata such as year of publication and authors were added. One or more labels were assigned to each paper corresponding to the cost or benefit discussed (\textit{open coding}). These labels were subsequently clustered into themes (\textit{axial coding}). The assignment of themes was done by selecting a new theme whenever a corresponding label appeared, otherwise reusing an existing theme. In the case of the potential benefits, these themes mainly correspond to the industry domain or enterprise capability where the benefit is gained. For the potential costs and considerations, the themes represent the areas where potential challenges or negative impacts might arise in the adoption or usage of AI.
Regarding the assignment of costs or benefits to the dimensions of sustainability, the first two authors independently performed the labelling with a later comparison and resolution of disagreements.

\subsection{Interviews} 
We selected semi-structured interviews to gather additional results from an industry perspective. This gave us a basic structure via our interview guide, but also left the possibility open to ask follow-up questions and steer the discussion into promising directions. The data synthesized from the interviews was used to compare and validate relevant review findings. Three out of the six interview participants belonged to the primary organization, while the remaining three were external consultants from two different companies that worked with the main organization.

\subsubsection{Study Preparation} An interview preamble was compiled and shared beforehand to explain the process and ethical considerations of the interview. A presentation was made at the beginning to familiarize the interviewees with the topic of the interview and at the end to share findings. Additionally, an interview guide was prepared with the main questions to structure the interview and help get answers to the research questions.

\subsubsection{Interview Execution and Data Synthesis}
Six industry experts, three from the organization and three who work as consultants, were interviewed by the first author in English via MS Teams. The interviews began with a presentation using a slide set to brief about the research topic and clarify sustainability concepts to set the stage. Then, preliminary questions were asked that covered practitioners’ demographics and experiences with sustainability and AI, followed by the main interview questions related to potential sustainability benefits and costs of integrating or using AI in software systems. In the end, the most relevant findings from the rapid review were shared for feedback and discussion. Each interview lasted 45-60 minutes, with five out of six recorded and manually transcribed for qualitative analysis. Manual notes were taken for one participant who did not consent to recording. The first author performed the data synthesis similarly to the rapid review, i.e., labels were attached to highlight interesting passages and concepts to answer the RQs (\textit{open coding}). Labels generated from the rapid review were reused as much as possible, but new labels were generated in the absence of similarity. There was frequent merging of labels. The labels were then axially coded by reusing the themes identified from the rapid review.

\section{Results}
\label{sec:results}
This section details the results from the rapid review and interviews with the high-level clustering.
A total of 151 studies were included in the review: 139 through the search queries and 12 additional ones via snowballing. The year of publication varied from 2007-2023. The result set includes a varied range of papers from different domains like healthcare, manufacturing, etc.
Regarding the interviews, the demographics of the six participants are described in Table~\ref{Participant Demographics}, where C1 is the ID of our collaborating company and C2 and C3 are two different companies that work with the main organization. Four out of the six participants had more than 10 years of experience. Most of the participants were either working on AI- or ML-related projects or were involved in investigations around the usage of it in the organization. Furthermore, the interviewees showed awareness of sustainability concepts. The results of the rapid review and interview findings are described per RQ in the following subsections.

\begin{table}[ht]
\centering
\caption{Participant Demographics} 
\setlength\tabcolsep{2pt}
\footnotesize{
\begin{tabular}{|l|l|l|p{1cm}|} 
 \hline
\textbf{Person ID} 
& \textbf{Company ID}
& \textbf{Role}
& \textbf{YoE} \\
 \hline
  P1 & C1 & Product Owner \& Domain Expert & 11.5  \\
 \hline
 P2 & C1 & ML Engineer &  3.5 \\
 \hline
  P3  & C1 & Data Scientist  & 3 \\
 \hline
P4  & C2 & IT Development Engineer  & 13 \\
 \hline
 P5  & C3 & Lead Data Scientist  & 12\\
 \hline
 P6  & C3 & Client Partner - Growth \& Delivery  & 20\\
 \hline
 \multicolumn{2}{l}{\footnotesize YoE - Years of Experience}\\
\end{tabular}}
\label{Participant Demographics}
\end{table}

\subsection{Potential Sustainability Benefits of AI (RQ1)}

\begin{table}
\centering
\caption{Top 10 themes and their most repeated labels from rapid review for potential sustainability benefits with labels of more than 6 mentions highlighted in bold}
\label{Tab:Sustaianbility-benefit-most-mentioned}
\begin{tblr}{
  width = \linewidth,
  colspec = {Q[250]Q[438]},
  hlines,
  vlines,
}
\textbf{Theme and total number of mentions} & \textbf{Most mentioned labels}                                                                                \\
Healthcare (3)                                & {\textbf{Disease prediction} (15)\\Faster drug discovery (4)\\Improved interpretation of medical imagery (4)} \\
Energy management (19)                                & {\textbf{Enhance energy efficiency} (6)\\\textbf{Optimising energy consumption} (6)}                          \\
Manufacturing (15)                                & {\textbf{Eco-friendly material synthesis} (8)\\Improved behaviour prediction of materials (5)}                \\
Employee wellbeing (12)                                & {Enhanced decision-making (5)\\Reduce rote and dangerous work (4)}                                            \\
Operation management (12)                                & {Improved failure forecasting (4)\\Improved supply chain planning (4)}                                        \\
Agriculture (10)                                & \textbf{Improve farm management practices} (6)                                                                \\
Environment (10)                                & {Precise climate modelling (3)\\Sustainable land management (3)\\Water resources management (3)}              \\
Education (9)                                 & \textbf{Promote better learning outcomes} (6)                                                                 \\
Urban planning (9)                                 & Forecasting water quality (4)                                                                                 \\
Public administration (8)                                 & Improved policymaking (3)                                                                                    
\end{tblr}
\end{table}

\begin{figure*}[htp!]
    \centering
    \subcaptionbox{Potential sustainability benefits of AI (the positive aspects)\label{fig:sub1}}{
        \includegraphics[width=\textwidth]{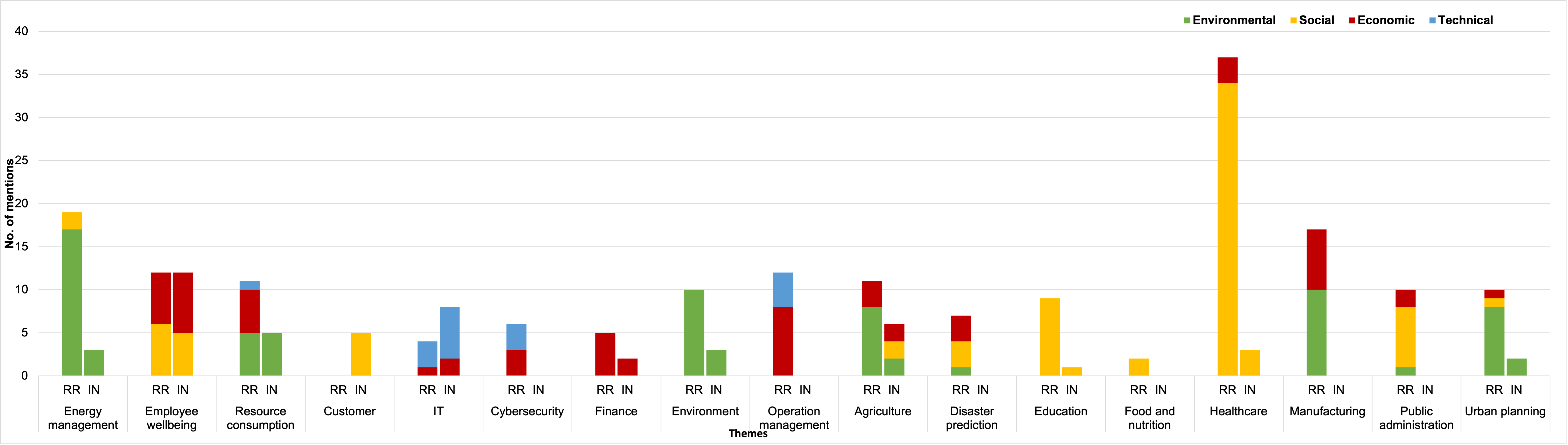}
        }
        \hfill
    \subcaptionbox{Potential sustainability costs of AI (the negative aspects)\label{fig:sub2}}{
        \includegraphics[width=0.6\textwidth]{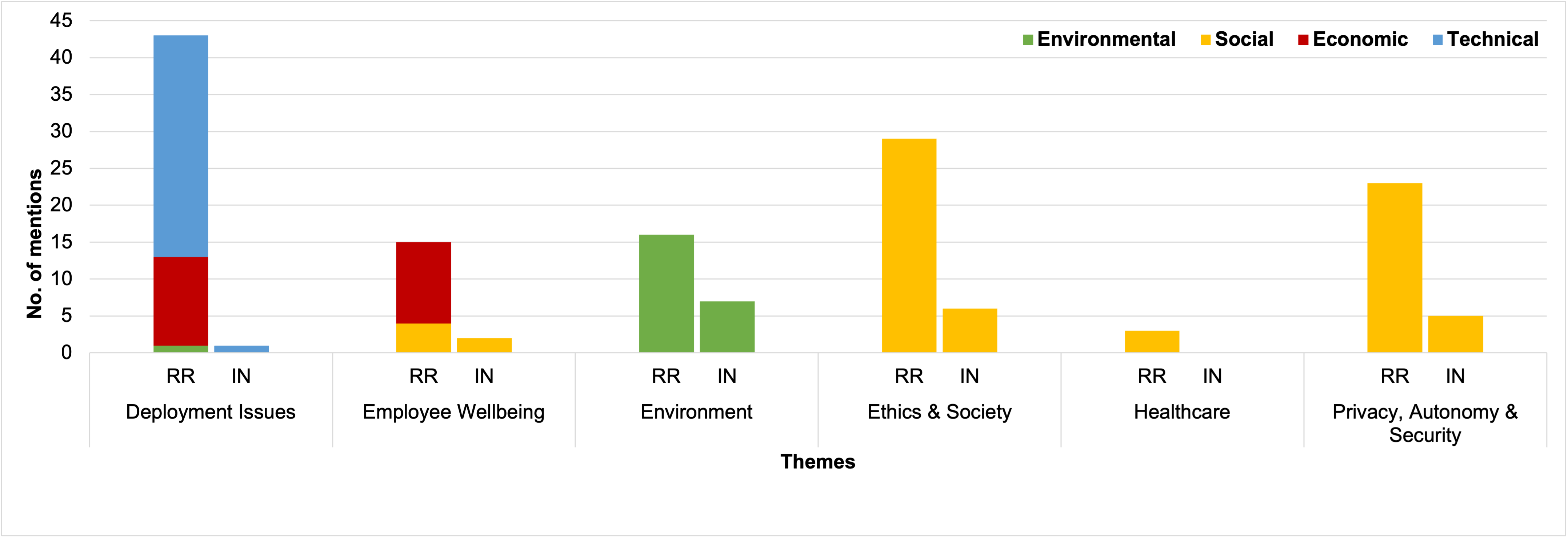}
    }
     \caption{Segmentation of themes based on dimensions of sustainability (RR - Rapid Review, IN - Interviews)}
    \label{fig:dimensions-grouped}
\end{figure*}

Table~\ref{Tab:Sustaianbility-benefit-most-mentioned} shows the total number of mentions of potential sustainability benefits related to the top 10 themes based on mentions from rapid review. The rapid review identified 16 high-level themes with 164 mentions distributed among 66 labels. The labels with at least six mentions are shown in bold in the table. An overview of the distribution of the impact of different themes among the different dimensions of sustainability is shown in Fig.~\ref{fig:sub1}.

The most prominent themes included \textit{healthcare}~(37), \textit{energy management}~(19), \textit{manufacturing}~(15), \textit{employee wellbeing}~(12) and \textit{operation management}~(12). From Fig.~\ref{fig:sub1}, we can deduce that \textit{healthcare}'s contribution to social sustainability was evident with advancements in disease prediction and medical imagery interpretation. Both the \textit{energy management} and \textit{manufacturing} themes predominantly affected the environmental dimension through benefits like enhanced energy efficiency~(6) and eco-friendly material synthesis~(8). \textit{Employee wellbeing} saw effects in both social and economic dimensions due to the nature of its concepts like enhanced decision-making and better work-life balance. In contrast, \textit{operation management} contributed to the economic and technical aspects of sustainability, with improved failure forecasting as the most mentioned label. The benefits related to \textit{information technology} and \textit{cybersecurity} were limited, appearing only four and three times, respectively.

On the other hand, in the case of the interviews, we found 11 high-level themes with 21 individual labels concerning potential sustainability benefits. The high-level themes of the benefits and their distribution are as follows: \textit{employee wellbeing}~(12), \textit{IT}~(8), \textit{resource consumption}~(5), \textit{customer}~(5), \textit{energy management}~(3), \textit{environment}~(3), \textit{healthcare}~(3), \textit{agriculture}~(2), \textit{finance}~(2), \textit{urban planning}~(2) and \textit{education}~(1). We also categorized each label based on its impact on different dimensions of sustainability. A visualization of it corresponding to each high-level theme is presented in Fig.~\ref{fig:sub1}.

Positives related to employees in the organization were prominent. Four labels were identified in this theme that could all be reused from literature findings, with AI's role in enhanced decision-making and work-life balance highlighted. For example, participant P1 talked about how AI could be leveraged to assist technical experts in optimising architectures for energy consumption. Three participants talked about how AI can reduce repetitive work by helping with automated summarization and transcription, repetitive work in SDLC and knowledge management respectively. Consequently, two participants mentioned that AI could contribute to an improved work-life balance and increased productivity. A noteworthy claim from participant P2 was that \enquote{The analysts are less on their computers, so they use less energy. You feel less work pressure. Therefore, you have a good workforce.}

Three labels related to \textit{IT} benefits were found, all of which could be reused from the review. Increased maintainability of software by generating comments and test cases was the sustainability benefit mentioned the most~(4), followed by intelligent automated quality assurance~(2) and faster time to market~(2). Participant P3 highlighted AI's significant potential in enhancing code quality and maintainability. They noted that AI could be used to automate the generation of comments and test cases, tasks that some developers often prefer not to undertake, which could lead to more efficient and higher-quality software development. 

In the case of the \textit{resource consumption} theme, one label was identified from the interviews which would be resued from the review findings. Five participants identified that AI could be used as a tool in their software development process to optimise resource utilization.

The \textit{customer} theme was created exclusively from the interview results. Two labels were revealed from the results. In one case, four out of six~(P1, P2, P4, P5) participants implied integrating AI in customer assistance of their products could help in better \enquote{personalization according to the age and demographics of the customer}. In another case, participant P6 mentioned that AI could assist in providing supplementary information for a particular context to the customers enabling them to take better decisions.

One potential sustainability benefit of using AI was mentioned by three participants: optimising energy consumption, which falls under the \textit{energy management} theme. Relating to the label \enquote{enhanced decision making}, participant P4 mentioned how AI could help in the organization's energy reduction targets by using productivity tools with green code suggestions from AI and by predicting the availability of green energy to do better distribution of workloads. A different participant~(P5) said that AI could help in defining an architecture that does more with less energy.

\begin{table}
\centering
\caption{Themes and their most repeated labels from rapid review for potential sustainability costs with labels of more than 6 mentions highlighted in bold}
\label{Tab:Categories-costs-most-mentioned}
\begin{tblr}{
  width = \linewidth,
  colspec = {Q[250]Q[438]},
  hlines,
  vlines,
}
\textbf{Theme and total number of mentions} & \textbf{Most mentioned labels}                                                                                                                                    \\
Deployment Issues (37)                               & {\textbf{Cost of training} (7),\\\textbf{Lack of generalizability due to overfitting} (7),\\\textbf{Trade-off data quality for data availability} (6)}            \\
Ethics  Society (29)                                & {\textbf{Imbibing of discriminatory characteristics from training data} (19),\\Fuzziness in ethical decision-making (4)}                                          \\
Privacy, Autonomy  Security (23)                                & {\textbf{Insufficient security of sensitive data} (8),\\\textbf{Profiling based on sensitive information} (6),\\Lack of transparency in data collection  use (5)} \\
Employee wellbeing (15)                                & {Increased Job displacement (5),\\Overdependence leading to decreased creativity (4)}                                                                             \\
Environment (16)                                & {\textbf{Increased energy consumption} (9),\\Increase of GHGs (3),\\Over-consumption of scarce raw materials (2),\\Disproportionate water consumption (2)}\\
Healthcare (3)          & Effectiveness skew by social group (2)                       
\end{tblr}
\end{table}

\subsection{Potential Sustainability Costs of AI (RQ2)}
Table~\ref{Tab:Categories-costs-most-mentioned} shows the distribution of mentions of themes identified from the results of the rapid review for potential sustainability costs of integrating AI. The rapid review for RQ2 highlighted 6 high-level themes with 123 mentions distributed among 32 labels. A visualization of the categorization of the sustainability dimensions across different labels in a high-level theme is shown in Fig.~\ref{fig:sub2}. 
 
The \textit{deployment issues}~(37) theme had the highest attention in the review, followed by \textit{ethics \& society}~(29), \textit{privacy, autonomy \& security}~(23), \textit{environment}~(16), \textit{employee wellbeing}~(15) and \textit{healthcare}~(3). The impact of labels related to \textit{deployment issues} from the review findings was mostly in the technical and economic dimensions, with focus on training costs~(7) or the lack of generalizability due to overfitting~(7). The social dimension was the target of the themes \textit{ethics \& society}, \textit{privacy, autonomy \& security} as well as \textit{healthcare}. This is because popular costs from the findings for \textit{ethics \& society} were imbibing discriminatory characteristics from training data~(19) and fuzziness in ethical decision-making~(4). Similarly, \textit{privacy, autonomy \& security} had insufficient security of sensitive data~(8) and profiling based on sensitive information~(6) as the major sustainability costs. The theme \textit{environment} included labels such as increased energy consumption~(9) while \textit{employee wellbeing} pointed to social and economic sustainability costs such as increased job displacement. 

The interview findings had 5 high-level themes with 13 unique labels, all of which could be reused from the review. The highest frequency of any individual label was four. Of these labels, eight were mentioned only once. The most repeated labels were of the themes \textit{environment}~(7), \textit{ethics \& society}~(6), followed by \textit{privacy, autonomy \& security}~(5), \textit{employee well-being}~(2) and \textit{deployment issues}~(1). Fig.~\ref{fig:sub2} shows the classification of themes based on their impacts on the dimensions of sustainability. 

In the \textit{ethics \& society} theme, the most mentioned cost was imbibing discriminatory characteristics from training data (19). Another cost in this area was reduced trustworthiness~(2) and increased distribution of misinformation~(1). All three labels were reused from the review.

In case of \textit{privacy, autonomy \& Security}, three costs were mentioned out of which two of them could be reused from the rapid review. Two participants raised concerns about fuzzy data ownership and insufficient security of sensitive data. Another potential cost mentioned once was the non-beneficial outcomes of using AI due to a lack of empathy or emotional intelligence. Adding to this cost, participant P6 mentioned recruitment companies using AI to screen people and do the first round of interviews and that AI was incapable of understanding the background of the person.

Costs related to the \textit{environment} theme synthesized from the interview were four labels, all of which were reused from the review. Four participants said increased energy consumption was a big sustainability cost related to integrating AI into the organization. Participant P4 highlighted the increased carbon footprint (GHG emissions) and the increased water footprint of cooling the data centres where AI is trained. Participant P3 also discussed the increased consumption of scarce raw materials to build supporting hardware.

Two unique labels were reported for the \textit{employee wellbeing} theme, both of which could be reused from the review. Only one participant~(P1) among the six saw job insecurity and job displacement as a potential sustainability cost. The participant further explained: \enquote{I think lower-level programming tasks that used to go to more junior people - I think partly that can be replaced by AI.}

For the theme of \textit{deployment issues}, a new label was created. One participant~(P3) stated that AI tended to deplete its data sources by continuous usage. This occurs since AI diverts away interaction from its data source, consequently reducing the data available to be trained on, e.g., a reduction in questions and answers on StackOverflow due to the increased usage of tools like ChatGPT. This could lead to reduced capabilities of AI over time.

\section{Discussion}
In this section, we interpret the literature and interview results concerning the potential sustainability-related trade-offs of integrating AI. To ensure the reliability of the outcomes, the most relevant findings were presented and discussed with our company partner. This helped combine academic research with practical insights from industry experts and ensured the usefulness and reliability of the study.

\subsubsection{Reflections on RQ1}
From the literature findings, some themes of benefits were not specifically useful for a typical IT organisation, e.g., \textit{agriculture} or \textit{healthcare}. The most mentioned relevant benefits were related to \textit{energy management}, \textit{employee wellbeing} and \textit{resource consumption}, emerging as commonly occurring themes in both the rapid review and interviews. Many of the synthesized benefits lie in the well-being of the organization’s employees, customers and ultimately enhancing their role as more sustainable and responsible contributors in the industry.

It is noteworthy that our primary studies were more skewed towards the general benefit of using AI in different domains and not specifically related to an IT organization focused on customer business. That might explain why some prominent themes in the review results were not mentioned in the interviews that frequently. On the other hand, IT-related benefits had several mentions in interview results, emphasizing aspects such as maintainability and automated quality assurance. These were also mentioned in the literature, although their occurrence was much less frequent. This highlights the differential focus between academia and industry in considering the sustainability benefits of AI.

Moreover, from the dimension distribution from Fig.~\ref{fig:sub1}, we see that only 4 out of the 16 themes in the review had an impact in only one dimension, as opposed to the interviews, which had 8 out of 11 themes having an impact only in one dimension. Thus, the review covered a broader spectrum across sustainability dimensions, whereas the interview findings were more focused, with most themes impacting a single dimension.

\subsubsection{Reflections on RQ2}
Contrary to RQ1 where the rapid review results for sustainability benefits had many themes not relevant for a financial organization, most of the findings from the rapid review on sustainability costs appear suitable to be evaluated as a trade-off when considering the use or integration of AI in the context of the organization. Interview findings highlighted concerns among AI practitioners about the potential environmental costs of AI to make sure that they still achieve their sustainability targets related to carbon footprint and energy reduction. 

Practitioners also voiced concerns around \textit{privacy, autonomy \& security} and \textit{ethics \& society}. One interesting cost that did not come up in the literature but was mentioned by a practitioner is AI’s nature to cannibalise its data sources. This occurs when AI diverts away the traffic from its data source, thereby reducing the available data sources to be trained on. In addition, the performance of large language models is also shown to decline with time~\cite{del2023Are}. This is a significant hurdle for the technical sustainability dimension when adopting generative AI in an organization. An additional insight from the interviews not identified in the review is that AI's lack of empathy could lead to non-beneficial outcomes in scenarios where empathy is essential.

The dimension distribution in Fig.~\ref{fig:sub2} presents that for the review results, four out of six themes had a negative impact on one dimension, while the interview findings suggest that all themes predominantly influence just one dimension. All costs considered by the participants mostly concerned social and environmental dimensions, with limited attention to the technical dimension.

\subsubsection{Implications}
The interviews with AI practitioners gave a glimpse of their deliberate, privacy-first, and energy-conscious decision-making mindset, aligning with sustainable and responsible AI adoption. Concerns related to brand safety, budget, regulations and availability of skilled persons were also mentioned in the interview. This hints that non-sustainability-related requirements are oftentimes aligned to mitigate some of the negative impacts of AI.

Despite awareness related to sustainable and responsible AI, the lack of documentation or a framework for practitioners in the main organization makes it harder in quantifying, auditing and mitigating the potential costs. It is recommended that practitioners consider adopting frameworks for sustainable ML~\cite{wenninger2022sustainable}, ethical design of AI~\cite{peters2020responsible} or apply green tactics~\cite{Jarvenpaa2023synthesis} to aid their decision-making to adopt AI. Finally, IT teams are encouraged to assess and evaluate the potential sustainability trade-offs of integrating AI using toolkits such as the Sustainability Assessment Framework (SAF) Toolkit or the Sustainability Awareness Framework (SusAF) and commit to AI adoption based on the results.\footnote{\url{https://github.com/S2-group/SAF-Toolkit}}\footnote{\url{https://zenodo.org/records/7342575}}
This study serves as a starting point for studying industry-specific sustainability trade-offs of integrating AI by comparing and validating the results of the literature with interviews. This contributes to increasing awareness and transparency among practitioners when making decisions related to the adoption of AI from a sustainability perspective.

\section{Threats to Validity}
\subsubsection{Internal Validity}
We used a lightweight secondary study method for faster results. This might result in a compromise in the rigor of the study, leading to a threat to the internal validity of the study. Despite this, a systematic search strategy combined with interviews helped to reduce possible downsides of the rapid review. Regarding the interviews, one potential threat was that practitioners might exhibit only a positive outlook towards adopting AI. Confidentiality and anonymity were guaranteed to foster a comfortable environment where they could voice their perspectives freely. To mitigate the influence of the interviewer on the participants, care was taken to refrain from asking leading questions. Since only a single researcher was involved in the study execution, this could lead to subjective bias. Nonetheless, the impact of this should not be significant since the findings from the literature were discussed with other researchers in the research group. The categorization of labels by sustainability dimensions was performed by two researchers independently.

\subsubsection{External Validity}
The rapid review includes multiple trade-offs such as using a single search engine (Google Scholar), limited search queries and a stopping criterion of 250 papers per search query. Focused snowballing of up to 30 references and citations from relevant papers partially counteracted this, but we acknowledge the limited breadth of our review. Additionally, the number of interview participants was scoped to six. This might lead to sampling bias because of the small sample size and the possibility of their perspective being homogeneous. One mitigation was to have three of the six interview participants as consultants from other organizations who worked for the main organization. Still, problems generalizing the findings to different organizations could be expected, e.g., with non-financial institutions or smaller companies. Future research could mitigate this with a systematic literature review and multiple interviews across different domains of the organization.  

\subsubsection{Construct Validity}
Data collection in this research mainly involved text-based data extracted from academic literature and interview transcripts. Consideration was taken to formalise and explain the definitions of research concepts to the interview participants to guarantee shared understanding. In cases where participant answers were ambiguous, nudging was performed for clarification. Lastly, since only one researcher was involved in the study, there might be a potential interpretation bias. However, we consider the likelihood of these issues to be fairly low, especially since all results were extensively discussed in the research team.

\section{Conclusion}
Although the sustainability benefits and costs of integrating or using AI have been studied separately, there is no comprehensive analysis to raise awareness among practitioners. To address this gap, we conducted a rapid review of 151 papers followed by semi-structured interviews with 6 AI practitioners to validate the results of the literature. The key findings of the research include the fact that some themes of benefits from the literature had no intersection with the results of the interviews, whereas the potential costs themes were mostly intersecting. Practitioners should take these findings into account when deciding to integrate or use AI-based systems. Since the interview was conducted with only a limited number of participants and organizations, more research is necessary to validate the generalizability of the findings. This study thus provides awareness from a sustainability perspective by conceptualising AI-related trade-offs to support practitioners in the industry in decision-making. We share a focused evidence briefing for practitioners online.\footnote{\url{https://t.ly/HLhaV}}
Building on our preliminary findings, an extensive study could be conducted with a broader domain of organizations across different geographical locations to comprehensively analyse the sustainability trade-offs of adopting AI.

\section*{Acknowledgment}
We thank our partners at the Dutch financial organization for their continuous feedback on study design and results. We also thank our interviewees for their valuable time and inputs.

\bibliographystyle{IEEEtranN}
\bibliography{references}

\end{document}